\documentstyle{aipproc}

\begin{document}

\title{Constraining Post-Newtonian Parameters With\\
Gravitational Waves}
\author{James S. Graber$^*$}
\address{$^*$407 Seward Square SE, Washington, DC 20003-1113\\ 
jgraber@mailaps.org}

\maketitle

\begin{abstract}
We re-express gravitational wave results in terms of post-Newtonian
parameters. Using these expressions, and some simplifying assumptions, we
compute that in a favorable case, i.e. a ten solar-mass black hole spiraling
in to a $10^6$ solar-mass black hole, LISA observations will be able to constrain
at the 10\% level or better a single combination of post-post-Newtonian
parameters one order higher than those already constrained by solar system
evidence. This significant constraint will be possible even if the signal-to-noise 
level is so low that the signal can only be found by matched
filtering, and hence only deviations between alternate signal
interpretations of order one half cycle or more can be detected. 
\end{abstract}

\address{$^*$407 Seward Square SE\\ 
Washington, DC 20003}


We report here some early results of a study that tries to make an explicit connection 
between classical parameterized post-Newtonian (PPN) theory as developed by 
Nordtvedt, Will and others\cite{nw1,willbook,damourreview,will1}, 
and one or more of the several forms of PPN analysis now being used to predict 
gravitational waves, 
particularly for the case of inspiralling binary black holes\cite{thornereview,damourwaves}.  
Classical PPN theory was used to analyze solar system tests of general relativity (GR) and 
compare results predicted by GR to the results predicted by any (metric-based) 
alternate gravity theory (AGT).  Classical PPN theory, plus observational results, 
puts parametric constraints on the 
spherical geometry of a condensed 
body in any viable AGT. Our main goal is to make the connection 
between the PPN analysis of AGTs and the perturbative PPN calculation of gravitational 
waves, and to determine the extent to which an observed inspiral might be able to put  
stronger constraints on the  PPN representation of a compact object in any possible AGT.  
In particular, we are interested to see if any second order constraints would be imposed, 
and if so to find an explicit expression for them. It is known that very high order 
calculations are required to produce accurate templates for the matched filter detection of  
gravitational waves from binary inspirals\cite{finn1,cutler1}.  This fact suggests that strong constraints will 
be placed on AGTs by the observation of any inspiral, but the inverse problem of 
deriving the constraints from the observed inspiral is not easy.  This problem is clearly 
related to the previously studied problem (often called parameter estimation) 
of recovering the parameters describing the 
inspiralling binary and the compact objects which it comprises from the observed 
gravitational wave information\cite{ryan1,ryan2,poisson2,poisson1,ft1}. Our study  
tries to extend parameter estimation to include determining the second order 
(or higher) PPN constraints 
placed on AGTs by an observed binary inspiral.  We report here some successful 
steps toward this goal. 

At the current stage of this study, we consider only the second-order non-rotating 
spherical solution of fully conservative metric alternate gravity theories.  
(If this is successful, we may  
attempt to extend the study to include rotating solutions and/or higher order terms.)  For 
this simplest case, we extend the classical PPN parameters to the second order by adding 
the second-order parameters $\delta$ and $\epsilon$ to the classical first-order parameters 
$\beta$ and $\gamma$ (sometimes called the Eddington parameters\cite{eddingtonbook}), 
resulting in the metric $ds^2 = 
   -TT(r)\,{dt}^2 + 
    SS(r)\,{dr}^2 + 
    SS(r)\,r^2\,{d\Omega }^2$
 where $TT(r) = 
   1 - \frac{2\,\alpha \,M}{r} + 
    \frac{2\,\beta \,M^2}{r^2} + 
    \frac{\frac{3}{2}\,\epsilon \,M^3}{r^3} + 
    \cdots$
and $SS(r) = 
   1 + \frac{2\,\gamma \,M}{r} + 
    \frac{\frac{3}{2}\,\delta \,M^2}{r^2} + 
    \cdots$. (Hereafter M is omitted, and $1/r$ is to be 
interpretted as $M/r$.)
Setting all the lowercase Greek parameters equal to one in the above metric recovers the 
leading terms in the 
{\it isotropic} Schwarzschild metric. This is the normalization used in 
classical PPN theory\cite{willbook}.  
We immediately convert to the  {\it standard} Schwarzschild metric form, which is used 
in gravitational wave research. The resulting metric is  $ds^2 = 
   -{T(r)}^2\,{dt}^2 + 
    {S(r)}^2\,{dr}^2 + 
    r^2\,{d\Omega }^2$, where $T(r) = 
   1 - \frac{\alpha }{r} + 
    \frac{-{\alpha }^2 + 2\,\beta  - 2\,\alpha \,\gamma }
     {2\,r^2} + \frac{\frac{3}{2}\,\alpha \,
        \left( -{\alpha }^2 + 2\,\beta  - 
          2\,\alpha \,\gamma  \right)  + 
       \frac{3}{2}\,\left( 8\,\beta \,\gamma  - 
          2\,\alpha \,{\gamma }^2 - 3\,\alpha \,\delta  - 
          3\,\epsilon  \right) }{3\,r^3} + 
    \cdots$, and 
$S(r) = 
 1 + \frac{\gamma }{r} + \frac{3\,\delta }{2\,r^2} +
    \cdots$ after conversion from the isotropic metric above.  Earlier works that considered 
similar second order extensions of PPN theory include Epstein and Shapiro\cite{epstein1}, 
Fischbach and Freeman\cite{fischbach1}, Richter and Matzner\cite{matzner1}, and
Damour and Esposito-Farese\cite{damour1}.

Our study has not progressed as far in the area of wave generation as in the area of 
spacetime geometry.  The second order PPN wave generation parameters have not yet been 
generalized analytically.  Therefore, for the purpose of computing the 
numerical results, the wave generation is assumed to be exactly identical in GR and the 
AGT tested in the numeric models.  This is unlikely to be true, and 
errs on the side of making this AGT too close to 
GR.  In the analytical calculations, the alternate wave generation variables are treated as 
totally free parameters (denoted WGn for wave generation of order n): this also is 
unrealistic, but it does allow for the parameterization of arbitrarily different ATGs.  

In this paper, we consider only phase information, and do not consider 
amplitude information, wave 
shape information or any other possible observable source of information about the 
inspiralling system.  This means that only one combination of spacetime parameters and 
wave generation parameters is constrained from the observable frequency results 
by our current analysis.  
The second-order PPN component of this constrained combination is the factor 
we wish to explicitly specify in terms of PPN parameters.  Then we wish to analyze 
the constraints it is possible to place on this parameter combination from a possible future
observation of a favorable binary inspiral by LISA.  
Strictly speaking, this constraint to the second-order parameters only applies if we assume two 
additional conditions: First that all lower order components are identical.  
Previous experimental constraints require this to be almost true for all viable theories.  
(Classical solar system first-order PPN tests constrain the first-order geometric parameters 
to be less than 1\% different from GR.  Binary pulsar spin down results require the 
leading order wave generation parameters to be near GR values.  In the specific AGT we 
test later in the numerical model section of this paper, the first condition is exactly true.
The second condition is that parameters of order higher than 2 are either also identical, 
or are too small to affect the results.  In general, this condition is open to question, 
and will be 
tested further in later work. In the highly simplified model we consider below, 
this second condition is also met.)   

Expressed most 
generally, this constrainable combination of second order geometric and wave generation 
parameters is 
$13725 + 2016\,WG2 - 
   2352\,{\beta }^2 + 16762\,\gamma  + 
   10752\,{\gamma }^2 - 
   2\,\beta \,\left( 8381 + 23856\,\gamma  \right)  + 
   19656\,\delta  + 19656\,\epsilon $.  But if we assume that the AGT 
agrees with GR in all orders lower than 2, then we set 
$\alpha  = \beta  = \gamma  = 1$ and the 
constrained parameter combination reduces to $-2843 + 224\,WG2 + 
   2184\,\delta  + 2184\,\epsilon $. If we further replace the second-order 
wave generation parameter WG2 
with its GR value of
$\frac{44711}{9072}$ the constrained combination reduces to 
$-20123 + 25272\,\delta + 25272\,\epsilon$. 
If we further replace $\delta$ and $\epsilon$ first with their GR values $\delta = 1$ 
and $\epsilon=1$ and then replace these parameters with their values in the AGT we have 
chosen to test, (which are $\delta = \frac{3}{2}$ and $\epsilon=\frac{8}{9}$) and then take 
the ratio, we find $\frac{30421}{40249}$.
  This is a difference of approximately 25\%.  The numeric results of our study indicate 
that this difference 
is detectable in the gravity wave frequency information from a binary inspiral of a type 
that LISA is reasonably likely to see\cite{ft1}.
 
The theory of gravitational waves is now well developed after intensive work in recent 
years.  Of the several alternate methods that have been employed, we followed the 
pathway pioneered by 
Teukolsky\cite{teuk1}, Poisson\cite{poisson3}, and Sasaki and Nakamura\cite{sn1}. 
This mathematical pathway now also incorporates much work done
both earlier and later by other authors including Regge and Wheeler\cite{rw1}, 
Newman and Penrose\cite{np1}, 
Price\cite{price1}, Bardeen\cite{bp1}, Press\cite{teuk2}, Chandrasekhar\cite{chandra}, 
and Tagoshi and Sasaki\cite{ts1,s1}. We followed this chain of papers, and 
tried to generalize the 
results by replacing GR formulas with their PPN equivalents. This plan has not yet been 
fully completed: to date, only certain of the formulas have been generalized.  We are thus 
able to proceed only part of the way analytically, after which we must switch to an 
at least partly numeric 
approach.
As indicated above, where the generalization is not yet available, we proceed in two ways: 
First, by adding an unknown parameter, and second,  by using the equivalent 
GR formula or result unchanged.  The major area where this has been necessary 
has been in the wave 
generation formula, where we use the WGn parameters described above.  These 
parameters can be replaced by PPN-based formulas when they become available, and 
replaced by the GR values for numerical work.  In the outline below we use mainly the 
terminology of Teukolsky\cite{teuk1}, hereafter Teukolsky1973, Poisson\cite{poisson4} 
hereafter Poisson1995,and then switch over to Finn and Thorne\cite{ft1} hereafter 
FinnThorne2000 for the last two 
functions which are the focus of the numerical work.

Briefly, the major steps in the above pathway are to characterize the geometry  
by dE/df, the rate of change in energy with respect to change in frequency
(which is a nonlinear but observable fiduciary for radius).  The wave 
generation process is characterized by the change in energy emitted as a 
function of time, dE/dt. Having these two factors allows 
one to compute the observable change in frequency with time, df/dt. 
(This relationship illustrates why it is impossible to 
disentangle geometry and wave generation without observing or 
measuring the development over time of at least one further 
parameter in addition to frequency).  
It is also necessary to  
determine the innermost stable circular orbit (ISCO).  By integrations one can compute the 
number of orbits to 
plunge N and the time to plunge T.  In the GR case, the lowest order  functions 
can be computed from the 
linearized theory and include the well known  quadrupole formula. The higher order 
corrections are 
expressed as a Taylor series in mass over radius or velocity over c, which are 
directly related for circular 
orbits.   The energy frequency function, dE/df is a function of geometry and is 
known exactly in both the 
Kerr and Schwarzschild cases, as is the ISCO, so the relevant series can be expanded 
as far as necessary.  It 
has also been possible to obtain similar expansions in the generalized case.  
(The exact results can be 
expressed as an integrals, or as solutions to fairly simple differential equations.)  
The formulas to second PPN order, are given in 
Equations (1) and (2). Equation (1) is 
the generalised result  of this study.  Equation (2) is the GR result, taken from 
Poisson1995, equation (44).
  When the PPN parameters are all set to their GR values of 1, Equation (1) 
reduces to Equation (2).

\begin{eqnarray}
\frac{dE}{df} &=& 
   1 + \frac{-9 + 10\,\beta  - 10\,\gamma }{6\,r}\nonumber\\&& + 
    \frac{-729 + 284\,{\beta }^2 - 684\,\gamma  - 
       184\,{\gamma }^2 + 
       4\,\beta \,\left( 171 + 326\,\gamma  \right)  - 
       702\,\delta  - 702\,\epsilon }{72\,r^2}
\label{cuts0}
\end{eqnarray}

\begin{equation}
\frac{dE}{df} =  1  - \frac{3}{2\,r}- \frac{81}{8\,r^2}
\label{cuts1199}
\end{equation}

The wave generation function is not known in general, even in GR, but for the 
case of a small body 
orbiting a large body, or other small perturbation, the Teukolsky equation can be used.  It 
is converted to a Regge-Wheeler equation via the Chandrasekhar transform.  A further 
transform converts the Regge-Wheeler equation to the Sasaki-Nakamura equation.  
(Alternately, this can be done in a single step via a generalized Chandrasekhar 
transform.)  This equation is then solved by a form of Taylor series expansion .  This is 
difficult, but results to quite high orders are now available in GR.  We have been able 
to obtain 
an exact equation similar to the Teukolsky radial equation in terms of an arbitrary 
spherical metric for the generalized spherical 
case we are considering. We obtain this equation by exactly the same Newman-Penrose 
process used by Teukolsky, except we use a generalized null tetrad.  The null tetrad we 
use reduces to the null tetrad used by Teukolsky when the specific standard 
Schwarzschild coordinate functions are substituted for the generalized functions used 
here.  The decoupled but not separated equation that results for the non-rotating case is 
analogous to Equation (4.7) of Teukolsky1973. The separated radial equation for the 
spherical non-rotating case is displayed in Equations (3) and (4). This equation 
is the analog of 
Equation (4.9) 
of Teukolsky1973 which is displayed in Equation (5) with the specific angular momentum 
parameter(a) set to zero to reduce to the spherical case we consider, and the coupling 
constant $\lambda$ also omitted for simplicity.

\begin{eqnarray}
0&=&(r^4\,S(r)\,{T(r)}^2\,
     \frac{d^2\,R}{{dr}^2}\nonumber\\  &&+ 
    r^3\,T(r)\,\left( 3\,r\,T(r)\,S'(r) + 
        S(r)\,\left( -2\,T(r) + r\,T'(r) \right) 
        \right) \,\frac{dR}
      {dr}\nonumber\\ &&+ 
    \left(KK(r)\, \right) \,R 
\label{cuts126}
\end{eqnarray}

where

\begin{eqnarray}
KK(r) &=& 
   {S(r)}^3\,\left( r^4\,w^2 - 6\,M\,r\,{T(r)}^2 \right)
        + 4\,i \,r^3\,w\,{S(r)}^2\,
     \left( -T(r) + r\,T'(r) \right) \nonumber\\ && + 
    r^3\,{T(r)}^2\,\left( -S'(r) + 2\,r\,S''(r) \right)\nonumber\\ &&
        + r^3\,S(r)\,\left( -4\,r\,{T'(r)}^2 + 
       T(r)\,\left( 5\,T'(r) + 2\,r\,T''(r) \right) 
       \right) 
\label{cuts13}
\end{eqnarray}

\begin{equation}
r\,\left( -2\,M + r \right) \,
     \frac{d^2\,R}{{dr}^2} + 
    2\,\left( M - r \right) \,
     \frac{dR}{dr} + 
    \frac{-12\,i \,M\,r\,\omega  + 
     r^2\,\omega \,\left( 4\,i  - r\,\omega  \right) }
     {2\,M - r}
\,R = 0
\label{cuts11}
\end{equation}

Note that we here use the second gauge mentioned above, and that the coefficients have 
not yet been 
expanded into Taylor series. The separation method in the generalized non-rotating 
case is exactly the same 
as in the Kerr case solved by Teukolsky, and the separated angular equation that  
results in the 
generalised non-rotating case is exactly identical to the separated angular equation 
obtained by Teukolsky 
in the Kerr case (Equation (4.10) of Teukolsky1973).

 The rest of the analytic calculation for N the number of orbits to plunge and T the time to 
plunge proceeds by integrations of Taylor series, and is exact to the accuracy of the series 
in both the GR case and the generalized case. (At this point we switch to following the 
formulas and the notation in FinnThorne2000)    This is as far as we proceeded using the 
unknown parameter approach for the wave generation function dE/dt.
In the remainder of this study we used the exact GR result, taken from Poisson1995 Equation 
(41), displayed here as Equation (6).

\begin{equation}
\frac{dE}{dt} =   1 - \frac{1247}{336\,r} + 
    \frac{4\,\pi }{r^{\frac{3}{2}}} - 
    \frac{44711}{9072\,r^2}
\label{cuts1198}
\end{equation}

In doing a numeric study, it is also necessary to 
choose one (or more) specific AGT.  We chose the exponential metric $ds^2 = 
   \frac{-{dt}^2}{e^{\frac{2\,M}{r}}} + 
    e^{\frac{2\,M}{r}}\,
     \left( {dr}^2 + 
       {d\Omega }^2\,r^2 \right)$, which is  
part of several theories and has been studied by many authors, including 
Yilmaz\cite{yilmaz1,yilmaz2}, 
Rosen\cite{rosen1}, Kaniel and Itin\cite{kaniel1,itin1}, 
Muench, Gronwald and Hehl\cite{hehl1}, 
Watt and Misner\cite{misner1}, and Leiter and Robertson\cite{lr1}.  
We also used the specific numeric values 
appropriate for the AGT exponential metric or for the Schwarzschild metric or for the 
Kerr metric to replace the generalized PPN parameters used earlier.  (The exponential 
metric theories agree with GR exactly at first PPN order, and differ by 10 to 30\% in the 
second order PPN parameters.) Thus we ended up with  three sets of two equations: one set 
for each of the three metrics, with each set including an equation for T, time to plunge, 
and N, number of orbits to plunge.  Using these analytic functions with numerical 
coefficients, we numerically build and compare three inspiral models, 
called ALT, KERR, and SCHW. Based on the favorable possibilities 
for LISA (the proposed orbiting 
gravitational wave observatory) discussed in FinnThorne2000, where many cases 
result in over 10,000 orbits being observable and favorable cases result in over 
100,000 orbits being observable,
we chose to make our models 
have 10,000 orbits from the 
beginning to the 
end of the observed inspiral.  
The three models are displayed in Table 1.

\begin{table}[htbp]
\caption{Three closely matched binary inspiral models.}
\label{table1}
    \begin{tabular}{|l|r|r|r|r|r|r|r|r|r|}
\tableline
Model&
\multicolumn{9}{l|} {\quad Number of cycles remaining before plunge:
\tablenote{Number of cycles remaining before plunge in the 
Kerr(KERR) and Schwarzschild(SCHW) models, when the alternate(ALT) model has 
exactly the tabulated number of cycles remaining.}  } \\
\tableline
ALT\tablenote{The ALT model has a mass of 
1 arbitrary unit and no angular momentum. }
&10000.00&7000.00&3000.00&1000.00&300.00&100.00&30.00&10.00&0.00\\
KERR\tablenote{The KERR model has a mass of 1.012 
arbitrary units and a specific angular momentum of .063 per arbitrary mass unit.}
&10000.00&7000.28&3000.00&999.92&299.96&99.98&29.99&10.00&0.00\\
SCHW\tablenote{The SCHW model has a mass of 1.041 
arbitrary units and no angular momentum. }
&10000.00&7008.91&3009.61&1004.29&301.41&100.48&30.15&10.05&0.00\\
\tableline
    \end{tabular}
\end{table}

To test 
whether gravitational wave frequency (phase) information will allow us to distinguish 
these two alternate theories, (GR and the exponential metric theory), we try to match 
them as closely as possible.
First, an arbitrary ALT model was computed.  Since we could vary the angular momentum in the 
Kerr model to fit the ALT model, but not vice-versa, we froze the ALT model first.   Then, 
the closest possible Schwarzschild solution model(SCHW) and 
the closest 
possible Kerr solution model (KERR)were fitted to the resulting gravitational wave frequency 
(or phase) 
pattern from the ALT model.  (Many different SCHW and KERR models were computed to 
find the closest possible one.) The final SCHW model 
has the same 
number of cycles as the ALT model and is fitted to match the frequency pattern of the 
ALT model 
exactly at the 
beginning and the end of the observed inspiral. The SCHW model was fitted to match the 
ALT model 
by adjusting the mass of the Schwarzschild black hole.  The resulting fitted mass of 
the central body in the 
SCHW model is 1.041 times the mass of the central body in the ALT model.  The KERR 
model was 
fitted to the alternate model by adjusting both the mass and the angular momentum of 
the Kerr black hole.  
The KERR model was was fitted to match the ALT model exactly at three points: the 
beginning and  the 
end of the observed inspiral, and at one intermediate point (3000 orbits).  It 
fits the non-rotating ALT 
pattern much more closely than does the Schwarzschild model. The final KERR model central body 
has a mass 1.012 times the mass of the ALT model central body 
and a specific angular momentum of .063 per unit mass.  
The resulting 
matches and mismatches 
are displayed in Table 1.  The closest possible Schwarzschild match differed from the 
ALT model by almost 
ten full orbits.  This is easily detectable.  The closest Kerr model differed from the 
ALT model by slightly 
more than one quarter orbit.  This is marginally detectable.  This model supports the 
possibility of constraining a second-order combination of PPN parameters at the 
10 to 30\% level from only 10,000 orbits.  More extensive calculations using 
up to 100,000 orbits support the 
possibility of constraints substantially below the 10\% level.

It is too early to reach conclusions, but one conclusion toward which the preliminary 
results of this study 
point is that it may be possible to constrain a single combination of 
second-order PPN parameters at the 
10 to 30\% level or better by 
gravitational wave frequency data.  This is based on the assumption that a favorable 
case will be observed 
by LISA, the proposed orbiting gravitational wave observatory.

\end{document}